\documentclass[prd,aps,preprint,nofootinbib,amssymb,eqsecnum,showkeys]{revtex4}
\usepackage{verbatim,graphics,graphicx,color,slashed}
\usepackage{ulem} 
\usepackage{hyperref}
\usepackage{changepage}

\def\lsim{\mathrel {\vcenter {\baselineskip 0pt \kern 0pt
    \hbox{$<$} \kern 0pt \hbox{$\sim$} }}}
\def\gsim{\mathrel {\vcenter {\baselineskip 0pt \kern 0pt
    \hbox{$>$} \kern 0pt \hbox{$\sim$} }}}
\def\slashchar#1{\setbox0=\hbox{$#1$}           
 \dimen0=\wd0                                 
  \setbox1=\hbox{/} \dimen1=\wd1               
\ifdim\dimen0>\dimen1                        
  \rlap{\hbox to \dimen0{\hfil/\hfil}}      
  #1                                        
  \else                                        
 \rlap{\hbox to \dimen1{\hfil$#1$\hfil}}   
   /                                         
  \fi}                                         %
\def\cpto{\mathrel {\vcenter {\baselineskip 0pt \kern 0pt
    \hbox{$CP$} \kern 0pt \hbox{$\longrightarrow$} }}}
\def\cptof{\mathrel {\vcenter {\baselineskip 0pt \kern 0pt
    \hbox{$~CP$} \kern 0pt \hbox{$\longleftrightarrow$} }}}

\begin{document}

\baselineskip=15pt

\preprint{}

\title{U-spin analysis of  CP violation in $B^- $ decays \\into three charged light pseudoscalar mesons}

\author{Dong Xu${}^{1}$\footnote{xudong1104@gmail.com}}
\author{Guan-Nan Li$^{1}$\footnote{lgn198741@126.com}}
\author{Xiao-Gang He$^{1,2,3}$\footnote{hexg@phys.ntu.edu.tw}}
\affiliation{${}^{1}$INPAC, SKLPPC and Department of Physics,
Shanghai Jiao Tong University, Shanghai, China}
\affiliation{${}^{2}$Physics Division, National Center for
Theoretical Sciences, Department of Physics, National Tsing Hua
University, Hsinchu, Taiwan} \affiliation{${}^{3}$CTS, CASTS and
Department of Physics, National Taiwan University, Taipei, Taiwan}

\date{\today $\vphantom{\bigg|_{\bigg|}^|}$}

\date{\today}

\vskip 1cm
\begin{abstract}
We carry out a $U$-spin symmetry analysis for CP violation in $B^- $ decays into three light $\pi^-\pi^-\pi^+$, $\pi^- K^-K^+$, $K^-K^-K^+$ and $K^- \pi^-\pi^+$ mesons. We clarify some subtle points in constructing decay amplitudes with $U=0$ formed by the two negatively charged light mesons in the final states. $U$-spin conserving momentum independent and momentum dependent decay amplitudes,  and $U$-spin violating decay amplitudes due to quark mass difference are constructed. 
\end{abstract}

\pacs{PACS numbers: }

\maketitle

\section{Introduction}

In this paper we carry out a $U$-spin symmetry analysis for $B^-$ decays into three light pseudoscalar mesons. Here the light pseudoscalar means one of the $\pi^-$, $\pi^+$, $K^-$ or $K^+$ mesons. The branching ratios and $CP$ asymmetries, defined by $A_{CP}(f^+) = (\Gamma(B^-\to f^-) - \Gamma(B^+ \to f^+))/(\Gamma(B^-\to f^-) + \Gamma(B^+ \to f^+))$,  for these decays have been measured experimentally although some of them still have large error bars. 

The CP asymmetries measured for the two $\Delta S = -1$, $B^- \to K^- \pi^-\pi^+$ and $B^- \to K^-K^-K^+$  final states are~\cite{lhcb1}
\begin{eqnarray}
&&A_{CP}(K^+\pi^+\pi^-) = +0.032\pm 0.008(stat) \pm 0.004(syst)\pm 0.007(J/\psi K^+)\;,\nonumber\\
&&A_{CP}(K^+K^+K^-) = -0.043\pm 0.009(stat) \pm 0.003(syst)\pm 0.007(J/\psi K^+)\;,
\end{eqnarray}
which are $2.8\sigma$ and $3.7\sigma$ away from zero, respectively. Recently BaBar collaboration also reported their measurement~\cite{babar} of $A_{CP}(K^+K^+K^-) = -0.017^{+0.019}_{-0.014}\pm0.014$ which is consistent with the LHCb result within $1.1\sigma$.

The other two $\Delta S = 0$ CP asymmetries are given by~\cite{lhcb2}
\begin{eqnarray}
&&A_{CP}(\pi^+\pi^+\pi^-) = +0.017\pm 0.021(stat) \pm 0.009(syst)\pm 0.007(J/\psi K^+)\;,\nonumber\\
&&A_{CP}(K^+\pi^+\pi^-) = -0.141\pm 0.040(stat) \pm 0.018(syst)\pm
0.007(J/\psi K^+)\;.
\end{eqnarray}
The significances are 4.2$\sigma$ and 3.0$\sigma$, respectively.

The branching ratios for these decays have also been measured with~\cite{brs}
\begin{eqnarray}
&&Br(\pi^-\pi^-\pi^+) = (15.2\pm 1.4)\times 10^{-6}\;,\;\;\;\;\;Br(\pi^- K^-K^+) = (5.0\pm 0.7)\times 10^{-6}\;,\nonumber\\
&&Br(K^-K^-K^+) = (34.0\pm1.0)\times 10^{-6}\;,\;\;Br(K^- \pi^-\pi^+) = (51.0\pm 3.0)\times 10^{-6}\;.
\end{eqnarray}

These charged 3-body $B^-$ decays can provide new information about the SM and for strong interaction which determine the hadronic matrix elements for $B$ decays. 
Several analyses based on flavor $SU(3)$ symmetry both algebraic and diagrammatic approaches without resonant 
contributions\cite{gronau1,xlh,gronau2,london}, and dynamic models including resonant 
contributions~\cite{he3,fajfer,hycheng,yadong} have been carried out. Analysis based on $U$-spin symmetry has also been carried out~\cite{gronau-u}. 

In general there is an amplitude $A_0$ related to $U = 0$ state composed of $K^-$ and $\pi^-$ which contributes to $B^-\to K^-\pi^-\pi^+$ and $B^-\to \pi^-K^-K^+$, but not to $B^-\to K^-K^- K^+$ and $B^- \to \pi^-\pi^-\pi^+$. Naively, because the boson particle nature of $K^-$ and $\pi^-$, the $U=0$ combination formed from $K^-$ and $\pi^-$ is identically zero if the corresponding amplitude is a constant independent of the kinematic or other dynamic variables of the $K^-$ and $\pi^-$ states. This would lead to  predictions  away from experimental observation. Similar situation occurs in $SU(3)$ analysis~\cite{xlh}. If $A_0$ depends on the momentum carried by the particles involved, and it is anti-symmetry under exchange of the momenta of the two negatively charged $K^-$ and $\pi^-$, $A_0$ needs not to be zero and the problem can be resolved. However, how the amplitudes depend on the momenta are not known. In this paper, we study how to construct momentum independent amplitudes, momentum dependent amplitudes by taking derivatives in the particle states, and also $U$-spin violating amplitudes due to quark mass difference.

\section{$U$-spin symmetry and the decay amplitudes}

$U$-spin symmetry is  a global $SU(2)$ symmetry taking $d$ and $s$ as the two elements in the fundamental representation, that is, $(q_i) = (d, s)$ form a $U$-spin doublet in quark flavor space.  The $\pi^-$ and $K^-$ therefore transform as a doublet, $(M_i) = (\pi^-, K^-)$.  The complex conjugate $(M^i)=(M_i)^* = (\pi^+, K^+)$ can also be written as a
doublet with lower indices $(\tilde M_i) = \epsilon_{ij}(M_j)^* = (K^+, - \pi^+)$.  The $B^-$ is composed of a $b$-quark and a light $u$-quark and therefore $B^-$ is a $U$-spin singlet. Using notations familiar with spin analysis, we can write the $d$, $s$, $\pi^- (\pi^+)$, and $K^- (K^+)$, states as
\begin{eqnarray}
&&|d\rangle = |1/2,+1/2\rangle\;,\;\;\;\;\;\;\;\;\;\;|s\rangle = |1/2, -1/2\rangle\;,\nonumber\\
&&|\pi^-\rangle = |1/2, +1/2\rangle_{\pi^-}\;,\;\;\;\;|K^-\rangle = |1/2, -1/2\rangle_{K^-}\;,\\
&&|K^+\rangle = |1/2, +1/2\rangle_{K^+}\;,\;\;|\pi^+\rangle = -|1/2, -1/2\rangle_{\pi^+}\;.\nonumber
\end{eqnarray}

In the SM the effective Hamiltonian $H^q_{eff}$ responsible to $B^-$ decays into three charged mesons are given by~\cite{heff}
\begin{eqnarray}
 H^q_{eff} = {4 G_{F} \over \sqrt{2}} [V_{ub}V^{*}_{uq} (c_1 O_1 +c_2 O_2)
   - \sum_{i=3}^{12}(V_{ub}V^{*}_{uq}c_{i}^{uc} +V_{tb}V_{tq}^*
   c_i^{tc})O_{i}],
\end{eqnarray}
where $q$ can be $d$ or $s$.  The coefficients
$c_{1,2}$ and $c_i^{jk}=c_i^j-c_i^k$, with $j$ and $k$ indicating the internal quark,
are the Wilson Coefficients (WC).  The tree WCs are of order one with, $c_1=-0.31$, and $c_2 = 1.15$. The penguin
WCs are much smaller with the largest one $c_6$ to be $-0.05$. These
WC's have been evaluated by several groups~\cite{heff}. $V_{ij}$ are the KM matrix elements.
In the above the factor $V_{cb}V_{cq}^*$ has
been eliminated using the unitarity property of the KM matrix.

The operators $O_i$ are given by
\begin{eqnarray}
\begin{array}{ll}
O_1=(\bar q_i u_j)_{V-A}(\bar u_i b_j)_{V-A}\;, &
O_2=(\bar q u)_{V-A}(\bar u b)_{V-A}\;,\\
O_{3,5}=(\bar q b)_{V-A} \sum _{q'} (\bar q' q')_{V \mp A}\;,&
O_{4,6}=(\bar q_i b_j)_{V-A} \sum _{q'} (\bar q'_j q'_i)_{V \mp A}\;,\\
O_{7,9}={ 3 \over 2} (\bar q b)_{V-A} \sum _{q'} e_{q'} (\bar q' q')_{V \pm A}\;,\hspace{0.3in} &
O_{8,10}={ 3 \over 2} (\bar q_i b_j)_{V-A} \sum _{q'} e_{q'} (\bar q'_j q'_i)_{V \pm A}\;,\\
O_{11}={g_s\over 16\pi^2}\bar q \sigma_{\mu\nu} G^{\mu\nu} (1+\gamma_5)b\;,&
O_{12}={Q_b e\over 16\pi^2}\bar q \sigma_{\mu\nu} F^{\mu\nu} (1+\gamma_5)b.
\end{array}
\end{eqnarray}
where $(\bar a b)_{V-A} = \bar a \gamma_\mu (1-\gamma_5) b$, $G^{\mu\nu}$ and
$F^{\mu\nu}$ are the field strengths of the gluon and photon, respectively.

For $\Delta S = -1$ and $\Delta S = 0$ $B^-$ decays, the quark $q$ are $s$ and $d$, respectively. The effective Hamiltonian
$H^q_{eff}$ has a simple $U$-spin structure for both the tree and penguin contributions and transforms as components in a doublet. 
It annihilates $B^-$ state and creates a final state $|1/2,+1/2>$ for $q=d$ (a $|1/2, -1/2>$ state for $q=s$). 

At the hadron level, the decay amplitude can be generically written as
\begin{eqnarray}
A  = \langle final\;\;state |H^q_{eff}|B^-\rangle = V_{ub}V^*_{uq} T(q) + V_{tb}V^*_{tq}P(q)\;,
\end{eqnarray}
where $T(q)$ contains contributions from the $tree$ as well as $penguin$ due to charm and up
quark loop corrections to the matrix elements,
while $P(q)$ contains contributions purely from
one loop $penguin$ contributions.

Since $H^q_{eff}$ annihilates $B^-$ state and creates a $U$-spin final state $|1/2,+1/2\rangle$ for $q=d$ (a $|1/2, -1/2\rangle$ state for $q=s$), only those final three meson states decayed from $B^-$ with correct $U$-spin quantum numbers  will be created. Therefore one needs to single out appropriate $|1/2,+1/2\rangle$ and
$|1/2-1/2\rangle$ ones formed by the three light meson in the final state to identify which combinations are allowed. The final three charged mesons can form $U$-spin eigen-states given in the following,
\begin{eqnarray}
|K^-K^-K^+\rangle &&= |1/2, -1/2\rangle_{K^-}|1/2, -1/2\rangle_{K^-}|1/2, +1/2\rangle_{K^+}\nonumber\\
&&=|1,-1\rangle_{K^-K^-}|1/2,+1/2\rangle_{K^+}\nonumber\\
&&= {1\over \sqrt{3}}|3/2,-1/2\rangle_1 - \sqrt{{2\over 3}}|1/2,-1/2\rangle_1\;,\nonumber\\
|\pi^-\pi^-\pi^+\rangle\;\;\;&&= -|1/2, +1/2\rangle_{\pi^-}|1/2, +1/2\rangle_{\pi^-}|1/2, -1/2\rangle_{\pi^+}\nonumber\\
&&=-|1,+1\rangle_{\pi^-\pi^-}|1/2,-1/2\rangle_{\pi^+}\nonumber\\
&&= -{1\over \sqrt{3}}|3/2,+1/2\rangle_1 - \sqrt{{2\over 3}}|1/2,+1/2\rangle_1\;,\\
|K^-\pi^-\pi^+\rangle\;\; &&= -|1/2, -1/2\rangle_{K^-}|1/2, +1/2\rangle_{\pi^-}|1/2, -1/2\rangle_{\pi^+}\nonumber\\
&&=-\left ({1\over \sqrt{2}}|1, 0\rangle_{K^-\pi^-} - {1\over \sqrt{2}}|0,0\rangle_{K^-\pi^-}\right )|1/2,-1/2\rangle_{\pi^+}\nonumber\\
&&= -{1\over \sqrt{3}}|3/2,-1/2\rangle_1 - {1\over \sqrt{6}}|1/2,-1/2\rangle_1+ {1\over \sqrt{2}}|1/2,-1/2\rangle_0\;,\nonumber\\
|\pi^-K^-K^+\rangle\;  &&= |1/2, +1/2\rangle_{\pi^-}|1/2, -1/2\rangle_{K^-}|1/2, +1/2\rangle_{K^+}\nonumber\\
&&=\left ({1\over \sqrt{2}}|1, 0\rangle_{\pi^-K^-}+ {1\over \sqrt{2}}|0,0\rangle_{\pi^-K^-}\right )|1/2,+1/2\rangle_{K^+}\nonumber\\
&&= {1\over \sqrt{3}}|3/2,+1/2\rangle_1 - {1\over \sqrt{6}}|1/2,+1/2\rangle_1 +{1\over \sqrt{2}}|1/2,+1/2\rangle_0\;,\nonumber
\end{eqnarray}
where the sub-indices ``0'' and ``1'' indicate the $U$-spin formed by the two negatively charged mesons.

Inspection of the above three charged meson final states one clearly sees that there are indeed $U$-spin  $|1/2, +1/2\rangle$ and $|1/2,-1/2\rangle$ eigen-states and  they will be the allowed final states. Indicating the strength for the $U=0$ and $U=1$,  formed by the two negatively charged mesons, by $A^T_0$ and $A^T_1$, one can write the $T$ amplitudes as~\cite{gronau-u}
\begin{eqnarray}
&&T(K^-(p_1)K^-(p_2)K^+(p_3)) = 2A^T_1(p_1,p_2,p_3)\;,\nonumber\\
&&T(\pi^-(p_1)\pi^-(p_2)\pi^+(p_3)) = 2A^T_1(p_1,p_2,p_3)\;,\nonumber\\
&&T(K^-(p_1)\pi^-(p_2)\pi^+(p_3)) = A^T_1(p_1,p_2,p_3) - A^T_0(p_1,p_2,p_3)\;,
\nonumber\\
&&T(\pi^-(p_1)K^-(p_2)K^+(p_3)) = A^T_1(p_1,p_2,p_3) - A^T_0(p_1,p_2,p_3)\;.
\end{eqnarray}
Here we have worked with the convention for the amplitudes, involving two identical particles in the final states, that we symmetrize the amplitude first and then divide a factor of 2 when calculating the decay width.

In the above, we have explicitly written the $A^T_{0,1}$ amplitudes as functions of momentum $p_i$ of the final mesons. This is particularly important for the $A^T_0$ amplitude. This amplitude is from the $U$-spin equal to 0 state formed by the two negatively charged mesons. This is an anti-symmetric combination in exchanging of $K^-$ and $\pi^-$. Since they are bosons, if there is no momentum dependence, the combination $K^- \pi^- - \pi^- K^-$ is identically equal to zero and therefore $A^T_0$ vanishes.
For $A^T_1$ amplitude, it is not necessary to be zero even if no momentum dependent due to the fact that it is from the symmetric combination of $K^-$ and $\pi^-$. The above analysis also applies to the penguin amplitudes $A^P_{0,1}$.

If the amplitudes are indeed momentum independent, one would predict~\cite{xlh}
\begin{eqnarray}
&&Br(\pi^+\pi^+\pi^-) = 2Br(\pi^+ K^+K^-)\;,\;\;\;\;\;\;Br(K^+K^+K^-) = 2Br(K^+\pi^+\pi^-)\;,\nonumber\\
&&A_{CP}(\pi^+\pi^+\pi^-) = A_{CP}(\pi^+ K^+K^-)\;,\;\;A_{CP}(K^+K^+K^-) = A_{CP}(K^+\pi^+\pi^-)\;. \label{rl1}
\end{eqnarray}
and 
 \begin{eqnarray}
&&{A_{CP}(\pi^{+}K^{+}K^{-}) \over A_{CP}(K^{+}
 \pi^{+}\pi^{-})}=-{Br(K^{-}
 \pi^{-}\pi^{+}) \over Br(\pi^{-}K^{-}K^{+})}\;,\;\;{A_{CP} (\pi^{+} \pi^{+}\pi^{-}) \over A_{CP} (K^{+}
 K^{+}K^{-})}=-{Br(K^{-} K^{-}K^{+}) \over Br(\pi^{-}
 \pi^{-}\pi^{+})}\;,\nonumber\\
 &&{A_{CP}(\pi^{+}K^{+}K^{-}) \over A_{CP}(K^{+} K^{+}K^{-}) }=-{Br(K^{-} K^{-}K^{+})\over Br(\pi^{-}K^{-}K^{+})}\;,\;\;{A_{CP}(\pi^{+} \pi^{+}\pi^{-}) \over A_{CP}(K^{+}
 \pi^{+}\pi^{-})}=-{Br(K^{-} \pi^{-}\pi^{+})\over Br(\pi^{-}
 \pi^{-}\pi^{+})}\;. \label{cpa}
\end{eqnarray}

The above relations do not agree with experimental data shown earlier. Therefore, one needs to make modifications. One of the possibilities to have a non-zero $A_0$ is to have momentum dependence for the amplitudes.
With momentum dependence, $A^T_0(p_1,p_2,p_3)$ can be non-zero, if it satisfies
\begin{eqnarray}
A^T_0(p_1,p_2,p_3) = - A^T_0(p_2,p_1,p_3)\;. \label{anti}
\end{eqnarray}

For example, a term of the form $c [(\partial^\mu K^-(p_1)) \pi^- (p_2)- K^-(p_1)  (\partial^\mu \pi^-(p_2))] \partial_\mu \pi^+(p_3)$ is $U$-spin zero, but not zero. $c$ is a momentum independent constant. Such a term in the Lagrangian gives an amplitude
 \begin{eqnarray}
 c(p_1-p_2)\cdot p_3\;,
 \end{eqnarray}
satisfying the requirement of Eq.(\ref{anti}).

In general, how the amplitudes depend on the momentum is not known. It can come from contributions to the amplitude due to exchange of particles (resonant contributions), and may also come from derivatives on the particle fields in the initial and final states. In the following we discuss how to construct non-resonant momentum dependent amplitude by taking derivatives on the meson fields. 

\section{Lowest order momentum dependent  amplitudes}

Construction of momentum dependent ampltiudes without resonant contributions can be done in a systematic way by numbers of derivate taken. From Lorentz invariance requirement, the derivatives will have even powers. The lowest order terms have zero derivatives and the terms of next order have two powers in derivatives. The construction is basically to use $B^-$, $M_i$, $M^i$ (or equivalently $\tilde M_i$) with different powers of derivatives, and the effective Hamiltonian  $H^q_{eff}$ to form $U$-spin 
singlet. To this end we denote the doublet formed by the effective Hamiltonian by $H^i$ ( or equivalently $H_i = \epsilon_{ij}H^j$). $H^1=1$ and $H^2=1$ represent $\Delta S = 0$ and $\Delta S = -1$ interactions, respectively. We will use $M_i$, $ M^i$, $H^i$ and $B^-$, plus derivatives  as our building blocks.

For $B^-$ decays into three charged mesons, we need two $M_i$, and one $M^i$. There is only one non-derivative $U$-spin singlet which can be constructed. It is given by
\begin{eqnarray}
a_1^0 M_iM^i M_j H^j B^-\;,\label{lead0}
\end{eqnarray}
where $a_1^0$ is a constant. In the above $B^-$ is going in and the light charged mesons are going out.

Expanding the above, we obtain the momentum independent decay amplitudes
\begin{eqnarray}
&&T^0(K^-(p_1)K^-(p_2)K^+(p_3) )= 2a^0_1\;,\nonumber\\
&&T^0(\pi^-(p_1)\pi^-(p_2)\pi^+(p_3) )= 2a^0_1\;,\nonumber\\
&&T^0(K^-(p_1)\pi^-(p_2)\pi^+(p_3)) = a^0_1\;,
\nonumber\\
&&T^0(\pi^-(p_1)K^-(p_2)K^+(p_3) )= a^0_1\;.
\end{eqnarray}
In the above, for the first two terms, the factor of 2 comes from identical particle effect.

Note that there is no equivalent amplitude for $A_0$ given earlier. The construction of such a term comes from
$U$-spin singlet formed from two $M_i$. Without derivatives, the only singlet can be formed is
\begin{eqnarray}
\epsilon^{ij}M_i M_j\;.
\end{eqnarray}
It is identically equal to zero due to the boson particle nature of $K$ and $\pi$. This provides another way to understand why there is no  momentum independent $A_0$ amplitude.

To have a non-zero contribution for $A_0$ amplitude, we have to include momentum dependent contributions. 
With derivatives, it is possible to have $U$-spin singlet formed from two negatively charged mesons. It is given by
\begin{eqnarray}
\epsilon^{ij}M_i \partial_\mu M_j\;.
\end{eqnarray}

To construct  Lorentz and $U$-spin invariant terms, one needs to take another derivative on $ M^i$ or $B^-$. We can have two types of terms containing $\epsilon^{ij}M_i \partial_\mu M_j$,
\begin{eqnarray}
(i)&&\epsilon^{ij}M_i (\partial_\mu M_j)\epsilon_{kl}M^k H^l (\partial^\mu B^-)\;,\nonumber\\ 
(ii)&&\epsilon^{ij}M_i (\partial_\mu M_j)\epsilon_{kl}( \partial^\mu  M^k) H^l B^-\;.
\label{anti1}
\end{eqnarray}

With two derivatives, there are also other terms. One can obtain them by taking appropriated derivatives from Eq.({\ref{lead0}). They are given by
\begin{eqnarray}
&&(a)\;\;(\partial_\mu M_i)M^i M_j H^j (\partial^\mu B^-)\;,
\;\;\;\;(b)\;\;M_i(\partial_\mu M^i )M_j H^j(\partial^\mu B^-)\,,\nonumber\\
&&(c)\;\;M_i M^i  (\partial_\mu M_j) H^j (\partial^\mu B^-)\;,\;\;\;\;(d)(\partial_\mu M_i)(\partial^\mu M^i )M_j H^j  B^-\;,\nonumber\\
&&(e)\;\;M_i(\partial_\mu M^i )(\partial^\mu M_j) H^j  B^-\;,\;\;\;\;(f)\;\;M_i(\partial^\mu M^i) (\partial_\mu M_j) H^j B^-\;.\label{partial1}
\end{eqnarray}
Note that if the two derivatives are both taken on one field, $\partial^2M_i$, $\partial^2M^i$ or $\partial^2 B^-$, using equations of motion, they do not produce terms different than Eq.(\ref{lead0}) in the $U$-spin limit.

The terms in Eq.(\ref{anti1}) and Eq.(\ref{partial1}) are not all independent because
\begin{eqnarray}
\epsilon^{ij}\epsilon_{kl} = \delta^i_k \delta^j_l - \delta^i_l\delta^j_k\;.
\end{eqnarray}
The two terms in Eq.(\ref{anti1}) can be expressed as linear combinations of the terms in Eq.(\ref{partial1})
\begin{eqnarray}
(i) = (c)-(a)\;,\;\;\;\;(ii) =(f)-(e)\;.
\end{eqnarray}

To emphasis the contributions for $A_0$ amplitude, we use $(i)$, $(ii)$, $(b)$, $((a)+(c))/2$, $(d)$ and $((e)+(f))/2$ as independent ones for $\Delta S = -1$ procceses and label them as
\begin{eqnarray}
&&(1)\;\;(\partial_\mu B^-) \pi^+[(\partial^\mu K^-) \pi^- - K^-(\partial^\mu \pi^-)]\;, 
\;\;(2)\;\;(\partial_\mu B^-)  K^- [K^- (\partial^\mu K^+) +\pi^-(\partial^\mu \pi^+)]\;, \nonumber\\
&&(3)\;\;(\partial_\mu B^-) [K^- (\partial^\mu K^-) K^+ + {1\over 2}((\partial^\mu (K^-)\pi^- + K^- (\partial^\mu \pi^-))\pi^+) ]\;,\nonumber\\
&&(4)\;\;B^- (\partial_\mu \pi^+)[(\partial^\mu K^-) \pi^- - K^-(\partial^\mu \pi^-)]\;,\;\;(5)\;\;B^- \partial_\mu K^- [(\partial^\mu K^+) K^- + (\partial^\mu \pi^+)\pi^-]\;, \nonumber\\
&&(6)\;\;B^- [(\partial_\mu K^-)K^- (\partial^\mu K^+) +{1\over 2}((\partial_\mu K^-) \pi^- + K^- (\partial_\mu \pi^-)) \partial^\mu\pi^+)]
\;. \label{p-term}
\end{eqnarray}

Replacing $\partial^\mu$ by the corresponding momentum $p^\mu$, we express the two derivative contributions to the decay amplitudes as
\begin{eqnarray}
{1\over m_B^2}\left (\alpha_{1}(1)+\alpha_{2}(2)+\alpha_{3}(3)+
\alpha_{4}(4)+\alpha_{5}(5)+\alpha_{6}(6)\right )\;.
\end{eqnarray}
In the above, we have normalized the dimension of the coefficients $\alpha_i$ so that they are dimensionless.
Similarly, one can define the amplitude $P^p$ for the penguin contribution. Similar expressions also apply to the $\Delta S = 0$ amplitudes.

Replacing $\partial^\mu$ by momentum $p^\mu$ in the above expressions, we obtain the tree momentum dependent 
amplitude $T^p$
\begin{eqnarray}
&&T^p(K^{-}(p_{1})K^{-}(p_{2})K^{+}(p_{3}))\nonumber\\
&&={1\over 2 m_B^2}\left (2\alpha_{2}p_{B}\cdot
p_{3}+\alpha_{3}p_{B}\cdot(p_{1}+p_{2})+2\alpha_{5}p_{1}\cdot
p_{2}+\alpha_{6}(p_{1}+p_{2})\cdot p_{3}\right )\;,\nonumber\\
&&T^p(K^{-}(p_{1})\pi^{-}(p_{2})\pi^{+}(p_{3}))\nonumber\\
&&={1\over 2 m_B^2}\left (
2\alpha_{2}p_{B}\cdot
p_{3}+\alpha_{3}p_{B}\cdot(p_{1}+p_{2})+2\alpha_{5}p_{1}\cdot
p_{2}+\alpha_{6}(p_{1}+p_{2})\cdot p_{3}\right . \nonumber\\
&&\left .+2(\alpha_{1}p_{B}\cdot (p_{1} -p_{2}) + \alpha_{4}(p_{1}-p_2)\cdot
p_{3} )\right )\;,\\
&&T^p(\pi^{-}(p_{1})\pi^{-}(p_{2})\pi^{+}(p_{3}))\nonumber\\
&&={1\over 2 m_B^2}\left (2\alpha_{2}p_{B}\cdot
p_{3}+\alpha_{3}p_{B}\cdot(p_{1}+p_{2})+2\alpha_{5}p_{1}\cdot
p_{2}+\alpha_{6}(p_{1}+p_{2})\cdot p_{3}\right )\;,\nonumber\\
&&T^p(\pi^{-}(p_{1})K^{-}(p_{2})K^{+}(p_{3}))\nonumber\\
&&={1\over 2 m_B^2}\left (
2\alpha_{2}p_{B}\cdot
p_{3}+\alpha_{3}p_{B}\cdot(p_{1}+p_{2})+2\alpha_{5}p_{1}\cdot
p_{2}+\alpha_{6}(p_{1}+p_{2})\cdot p_{3}\right . \nonumber\\
&&\left .+2(\alpha_{1}p_{B}\cdot (p_{1} -p_{2}) + \alpha_{4}(p_{1}-p_2)\cdot
p_{3} )\right )\;.\nonumber \label{p-amp}
\end{eqnarray}
In the above, the terms $\alpha_{1, 4}$  and $\alpha_{2,3,5,6}$ contribute to $A_0$ and $A_1$  respectively.

Note that in the $U$-spin symmetric limit, one has
\begin{eqnarray}
&&T^p(K^{-}(p_{1})K^{-}(p_{2})K^{+}(p_{3})) = T^p(\pi^{-}(p_{1})\pi^{-}(p_{2})\pi^{+}(p_{3}))\;,\nonumber\\
&&T^p(K^{-}(p_{1})\pi^{-}(p_{2})\pi^{-}(p_{3}))=T^p(\pi^{-}(p_{1})K^{-}(p_{2})K^{+}(p_{3}))\;.
\end{eqnarray}
Similarly, one can write down the penguin amplitude $P^p$.

 Neglecting the masses of $K$, and $\pi$, we have:
\begin{eqnarray}
&&T^p(K^{-}(p_{1})K^{-}(p_{2})K^{+}(p_{3}))= T^p(\pi^{-}(p_{1})\pi^{-}(p_{2})\pi^{+}(p_{3}))\nonumber\\
&&= \frac{1}{2m^2_B}\left [ (s + t) (2\alpha_2-\alpha_3-2\alpha_5+\alpha_6 ) + 2m^2_B(\alpha_3 + \alpha_5)\right ]\;,
\nonumber\\
&&T^p(K^{-}(p_{1})\pi^{-}(p_{2})\pi^{+}(p_{3}))= T^p(\pi^{-}(p_{1})K^{-}(p_{2})K^{+}(p_{3}))\\
&&=\frac{1}{4m_B^2}\left [(s + t) (2\alpha_2-\alpha_3-2\alpha_5+\alpha_6 ) + 2m^2_B(\alpha_3 + \alpha_5) - 2(s-t)(\alpha_{1}+\alpha_{4}) \right ]\;,\nonumber \label{ppp11}
\end{eqnarray}
where $s=(p_{2}+p_{3})^2$ and $t=(p_{1}+p_{3})^2$.

\section{Leading $U$-spin symmetry breaking contributions}

$U$-spin symmetry is broken by quark mass difference which will modify the decay amplitudes. We now study how to obtain the leading amplitudes for $U$-spin violating amplitudes due to quark mass difference.
The mass matrix for $d$ and $s$ quarks is given by
\begin{eqnarray}
(\chi^i_j) = \left ( \begin{array}{cc}
m_d&0\\
0&m_s
\end{array}\right )
= {m_d+m_s\over 2}\left ( \begin{array}{cc}
1&0\\
0&1
\end{array}\right ) + {m_d - m_s\over 2}\left ( \begin{array}{cc}
1&0\\
0&-1
\end{array}\right )\;.
\end{eqnarray}
From the above, one sees that the mass matrix transforms as a linear combination of a $U$-spin singlet (the piece proportional to the unit matrix) and a triplet (the piece proportional to $\sigma_3$).

The construction of contributions due to quark masses can be obtained by inserting $\chi$ at appropriate places in Eq.(\ref{lead0}) and contracting the indices appropriately. The piece proportional to unit matrix will produce a decay amplitude proportional to the momentum independent $U$-spin amplitudes which can be absorbed into the momentum independent part. Only the term proportional to $\sigma_3 = (\sigma^i_j)$ term contains new information. We find two independent terms
\begin{eqnarray}
&&\beta_1M_i\sigma^i_j M^j M_k H^k B^- \;,\nonumber\\
&&\beta_2M_iM^i M_j\sigma^j_k H^k B^-\;. 
\end{eqnarray}

Expanding the above terms, we obtain the $U$-spin breaking contributions to the decay amplitudes
\begin{eqnarray}
&&T^b(K^-(p_1)K^-(p_2)K^+(p_3))= -2\beta_1 -2 \beta_2\;,\nonumber\\
&&T^b(\pi^-(p_1)\pi^-(p_2)\pi^+(p_3))= 2\beta_1 + 2 \beta_2\;,\nonumber\\
&&T^b(K^-(p_1)\pi^-(p_2)\pi^+(p_3)) = \beta_1 -\beta_2\;,
\nonumber\\
&&T^b(\pi^-(p_1)K^-(p_2)K^+(p_3))= -\beta_1+ \beta_2\;. \label{u-break}
\end{eqnarray}

We have
\begin{eqnarray}
T^b(K^{-}K^{-}K^{+})-T^b(K^{-}\pi^{-}\pi^{+})=T^b(\pi^-K^{-}K^{+})-T^b(\pi^{-}\pi^{-}\pi^{+})\;.
\label{relation2}
\end{eqnarray}

Note that the above $U$-spin breaking terms do not have breaking terms related to the $A_0$ amplitude which should be there in general~\cite{gronau-u}. 
This is because the fact that to have a non-zero $A_0$ amplitude, derivative terms must be involved as discussed earlier. Including derivative terms, one can write two terms
\begin{eqnarray}
(i)\;\gamma_1\partial_\mu (M_i) M_j \epsilon^{ij} \partial^\mu(M^k)\sigma_k^l H^m\epsilon_{lm} B^-\;,\;\;
(ii)\;\gamma_2\partial_\mu (M_i) M_j \epsilon^{ij} M^k\sigma_k^l H^m\epsilon_{lm} \partial^\mu B^-\;.
\end{eqnarray}
However, the above two terms are equivalent to each other, to first order in light quark mass, because the relation
\begin{eqnarray}
(i) + (ii) = \partial^\mu (\partial_\mu (M_i) M_j \epsilon^{ij} M^k\sigma_k^l H^m\epsilon_{lm} B^-)
- \partial^\mu (\partial_\mu (M_i) M_j \epsilon^{ij}) M^k\sigma_k^l H^m\epsilon_{lm} B^-\;.
\end{eqnarray}
The first term in the above is a total derivative term which does not play a role. The second term is proportional to $(m^2_K - m^2_\pi)$
which is one order higher in light quark mass expansion compared to the the leading terms proportional to $\beta_i$ and can be neglected. Therefore (i) and (ii) are equivalent. Let us use (i) for discussion, one obtains additional corrections $T^{bp}$ to $T^b$
with 
\begin{eqnarray}
&&T^{bp}(K^-(p_1)K^-(p_2)K^+(p_3) )= 0\;,\nonumber\\
&&T^{bp}(\pi^-(p_1)\pi^-(p_2)\pi^+(p_3) )= 0\;,\nonumber\\
&&T^{bp}(K^-(p_1)\pi^-(p_2)\pi^+(p_3)) = \gamma_1 (p_1-p_2)\cdot p_B\;,
\nonumber\\
&&T^{bp}(\pi^-(p_1)K^-(p_2)K^+(p_3) )= -\gamma_1 (p_1-p_2)\cdot p_B\;. \label{u-break}
\end{eqnarray}

For consistence one should also now include terms of the form, $\partial^2 M_i M^i M_j H^j B^-$, $M_i \partial^2 M^i M_j H^j B^-$ and $M_i M^i \partial^2 M_j H^j B^-$, because $\partial^2 K = m^2_K K$ and $\partial^2 \pi = m^2_\pi \pi$ (the difference in masses breaks $U$-spin). These terms have been which neglected in the $U$-spin limit. When $U$-spin breaking is considered, they should be included. However, when expanding these terms using equations of motion, all resulting terms can be absorbed into terms proportional to $\beta_i$. We not need to write them again.

\section{Conclusion and discussions}

In the previous sections, we have studied construction of decay amplitudes for  $B^-$ to three charged light pseudoscalar mesons from $U$-spin symmetry considerations. The construction discussed  has many things in common with flavor $SU(3)$ symmetry analysis for these decays. We conclude the paper by making a comparison with $SU(3)$ construction of the decay amplitudes and summarize the main numerical results. 

Flavor $SU(3)$ symmetry contains $U$-spin symmetry. Therefore one expects that the same form of decay amplitudes will result for the same initial and final particles. Indeed we find the corresponding contributions of these two analyses. 

The total decay amplitudes $T_t$ and $P_t$ can be written as
\begin{eqnarray}
T_t = T^0 +T^p + T^b\;,\;\;P_t = P^0 + P^p + P^b\;.
\end{eqnarray}
In the above, $T^b$ and $P^b$ include $T^{bp}$ and $P^{bp}$ also, respectively.

In the analysis of flavor $SU(3)$ symmetry in Ref.\cite{xlh}, the amplitudes were written as
\begin{eqnarray}
T_t = T + T^p + \Delta T\;,\;\;P_t = P + P^p + \Delta P\;.
\end{eqnarray}

Apart from the identical factor conventions difference here and that used in Ref.\cite{xlh}, 
the roles of $T^0$, $P^0$ and $T^p$ and $P^p$ here are played by $T$, $P$, $T^p$ and $P^p$ in Ref.\cite{xlh}, respectively.
In Ref.\cite{xlh}, $\Delta T$ and $\Delta P$ amplitudes look different than what have been defined here, since
$\Delta T(K^-K^-K^+)$ is not equal to  $-\Delta T(\pi^-\pi^-\pi^+)$, and $\Delta T(K^-\pi^-\pi^+)$ is not equal to  $-\Delta T(\pi^- K^- K^+)$ as should be here shown in Eq.\ref{u-break}. 
However, if one shifts the definitions of $T^0$ and $\Delta T$ amplitudes as 
\begin{eqnarray}
&&\tilde T^0(\pi^-\pi^-\pi^+)=\tilde T^0(K^-K^-K^+) =T^0(\pi^-\pi^-\pi^+)+  T^b(\pi^-\pi^-\pi^+)\;,\nonumber\\
&&\tilde T^0(K^-\pi^-\pi^+)=\tilde T^0(\pi^- K^- K^+) = T^0(K^-\pi^-\pi^+) + {1\over 2} T^p(\pi^-\pi^-\pi^+)\;,\nonumber\\
&&\Delta \tilde T( K^-K^- K^+)=  T^b( K^-K^- K^+) -  T^b(\pi^-\pi^-\pi^+)\;,\nonumber\\
&&\Delta \tilde T( K^-\pi^- \pi^+)=  T^b( K^-\pi^- \pi^+) - {1\over 2} T^b(\pi^-\pi^-\pi^+)\;,\nonumber\\
&&\Delta \tilde T( \pi^-K^- K^+)=  T^b( \pi^-K^- K^+) - {1\over 2} T^b(\pi^-\pi^-\pi^+)\;,
\end{eqnarray}
the amplitudes $\tilde T$ and $\Delta \tilde T$ are equivalent to $T'$ and $\Delta T'$ defined in  Ref.\cite{xlh}. 
In the above  the factor 1/2 is due to different convention of identical factor in amplitudes.
The $U$-spin symmetry and $SU(3)$ symmetry for $B^- $ decays into three charged pseudoscalar mesons are equivalent. 

As far as in obtaining the forms of decay amplitudes for $B^-$ decays into three charged light pseudoscalar mesons is concerned, the $U$-spin symmetry analysis is considerably simpler than that of the $SU(3)$ symmetry analysis. However, $SU(3)$ analysis can also apply to some of the final three pseudoscalar mesons to be neutral ones and also also include $B_d^0$ and $B_s$ decay into three psesudoscalar mesons. 

The numerical fitting to data for $B^-$ to three light pseudoscalar mesons will be the same in both approaches. We summarize below the main conclusions regarding numerical analysis with data obtained in Ref.\cite{xlh} in the $U$-spin language.

With just $U$-spin conserving momentum independent amplitudes $T^0$ and $P^0$, one would obtain relations given in Eqs.(\ref{rl1}) and (\ref{cpa}). The LHCb data shown in Section I obviously do not support the branching ratio relations given by Eq.(\ref{rl1}). The relations for CP asymmetry $A_{CP}$ given in Eqs.(\ref{rl1} and (\ref{cpa})
do not agree with data  neither, except the ratio $A_{CP} (\pi^{+} \pi^{+}\pi^{-}) / A_{CP} (K^{+}
 K^{+}K^{-})$. The LHCb data $A_{CP} (\pi^{+} \pi^{+}\pi^{-}) / A_{CP} (K^{+}
 K^{+}K^{-}) = -2.7\pm 0.9$ agrees with the predicted value~\cite{gronau1} $-2.2\pm 0.2$ very well using Eq.(\ref{cpa}).
If experimental data at the LHCb will be further confirmed, one needs to include contributions from beyond the  $U$-spin conserving momentum independent effects to explain the data. It may help if we
take the momentum dependent and $U$-spin breaking contributions in
consideration.

Adding $U$-spin conserving momentum dependent amplitudes $T^p$ and $P^p$, the degeneracy between the amplitudes for $A(\pi^-\pi^-\pi^+)$ and $A(\pi^- K^- K^+)$, and $A(K^-K^-K^+)$ and $A(K^-\pi^-\pi^+)$ can be lifted by a new piece of contribution, the term proportional to $s-t$ in Eq.(3.13). Because this new contribution does not interfere with the other contributions, if it enhances the branching ratios of $B^- \to K^- \pi^- \pi^+$, it also enhances $B^- \to \pi^-  K^- K^+$ compared with $B^- \to K^- K^- K^+$ and $B^- \to \pi^- \pi^- \pi^+$, respectively. This does not help to improve fit to data which requires enhancement of branching ratio for  $B^-\to K^- \pi^- \pi^+$, but reduction for $B^- \to \pi^- K^- K^+$. 

The experimental data can be explained  by including $U$-spin conserving $T^0$ and $P^0$, and $U$-spin breaking terms $T^{b} $ and $P^b$ if these terms are sizable compared with $T^0$ and $P^0$.   Without including $T^p$ and $P^p$ terms, data can already be explained. One may wonder what will happen if both momentum dependent and $U$-spin breaking terms are included, such as  whether one can have small $U$-spin breaking contribution and/or have large $T^p$ and $P^p$.  It has been shown in Ref.\cite{xlh}  that including both the momentum dependent and $U$-spin breaking contributions, one still cannot obtain small $U$-spin breaking amplitudes to explain data. However, one can find solutions with sizable $T^p$ and $P^p$ compared with $T^0$ and $P^0$.

\begin{acknowledgments}

XG would like to thank Hai-Yang Cheng and Jusak Tandean for useful comments.
The work was supported in part by MOE Academic Excellent Program (Grant No: 102R891505) and NSC of ROC, and in part by NNSF(Grant No:11175115) and Shanghai Science and Technology Commission (Grant No: 11DZ2260700) of PRC.

\end{acknowledgments}


\begin{thebibliography}{99}


 \bibitem{lhcb1} R. Aaij et al. (LHCb Collaboration), arXiv:1306.1246[hep-ex].

  \bibitem{babar} J.P. Lees et al. (BaBar Collaboration), arXiv:1305.4218[hep-ex].

 \bibitem{lhcb2} R. Aaij et al. (LHCb Collabortion), Report No. LHCb-CONF-2012-028, presented at the International conference on High Energy Physics, Melbourne, Australia, July 2012; R. Aaij et al. (LHCb Collabortion), arXiv:1310.4740[hep-ex].

 \bibitem{brs} Y. Amhis et. al., Heavy Flavor Average group, arXiv:1207.1158 and online update at http://www.slac.stanford.edu/xorg/hfag.

\bibitem{gronau1} B. Bhattacharya, M. Gronau and J. L. Rosner, arXiv:1306.2625.

\bibitem{xlh} D.~Xu, G.~-N.~Li and X.~-G.~He,
  arXiv:1307.7186 [hep-ph].
    
  \bibitem{gronau2}
  M.~Gronau, O.~F.~Hernandez, D.~London and J.~L.~Rosner,
  Phys.\ Rev.\ D {\bf 52}, 6356 (1995)
  [hep-ph/9504326].

  \bibitem{london}
  B.~Bhattacharya, M.~Imbeault and D.~London,
  arXiv:1303.0846 [hep-ph];
N.~R.~-L.~Lorier, M.~Imbeault and D.~London,
  Phys.\ Rev.\ D {\bf 84}, 034040 (2011)
  [arXiv:1011.4972 [hep-ph]];
M.~Imbeault, N.~R.~-L.~Lorier and D.~London,
  Phys.\ Rev.\ D {\bf 84}, 034041 (2011)
  [arXiv:1011.4973 [hep-ph]];
N.~Rey-Le Lorier and D.~London,
  Phys.\ Rev.\ D {\bf 85}, 016010 (2012)
  [arXiv:1109.0881 [hep-ph]].
  
  
  \bibitem{he3}
  N.~G.~Deshpande, G.~Eilam, X.~-G.~He and J.~Trampetic,
  Phys.\ Rev.\ D {\bf 52}, 5354 (1995)
  [hep-ph/9503273].

  \bibitem{fajfer}
  S.~Fajfer, T.~-N.~Pham and A.~Prapotnik,
  Phys.\ Rev.\ D {\bf 70}, 034033 (2004)
  [hep-ph/0405065].

  \bibitem{hycheng}
  H.~-Y.~Cheng and K.~-C.~Yang,
  Phys.\ Rev.\ D {\bf 66}, 054015 (2002)
  [hep-ph/0205133];
  H.~-Y.~Cheng, C.~-K.~Chua and A.~Soni,
  Phys.\ Rev.\ D {\bf 76}, 094006 (2007)
  [arXiv:0704.1049 [hep-ph]].


  \bibitem{yadong}
  Z.~-H.~Zhang, X.~-H.~Guo and Y.~-D.~Yang,
  Phys.\ Rev.\ D {\bf 87}, 076007 (2013)
  [arXiv:1303.3676 [hep-ph]];
  H.~-Y.~Cheng and C.~-K.~Chua,
  arXiv:1308.5139 [hep-ph].

  \bibitem{gronau-u}
  M.~Gronau,
  arXiv:1308.3448 [hep-ph].

  
  
  \bibitem{heff}
  G.~Buchalla, A.~J.~Buras and M.~E.~Lautenbacher,
  Rev.\ Mod.\ Phys.\  {\bf 68}, 1125 (1996);
  M.~Ciuchini, E.~Franco, G.~Martinelli and L.~Reina,
  Nucl.\ Phys.\ B {\bf 415}, 403 (1994);
  N.~G.~Deshpande and X.~-G.~He,
  Phys.\ Lett.\ B {\bf 336}, 471 (1994).




\end{thebibliography}
\end{document}